\documentclass[letter]{aa} % for the letters 
\usepackage{amsmath}
\usepackage{soul}
\usepackage{tablefootnote}
\usepackage{bm}
\usepackage{xcolor}
\usepackage[normalem]{ulem}
\usepackage[] {siunitx}
\DeclareSIUnit\angstrom{Å}
\DeclareSIUnit\gauss{G}
\DeclareSIUnit\Ic{I_c}
\definecolor{al}{rgb}{0.6,0.2,0.0}
\definecolor{as}{rgb}{0.8,0.2,0.3}
\definecolor{mvn}{rgb}{0.5,0.5,0.9}
\definecolor{fis}{rgb}{0.82,0.41,0.12}

\usepackage{graphicx}
\usepackage{txfonts}
\usepackage[colorlinks,linkcolor=blue,citecolor=blue,linktocpage=true,breaklinks, 
plainpages=false,urlcolor=blue]{hyperref}

\definecolor{iimm}{cmyk}{1.0,0.0,1.0,0.3}
\definecolor{ttkk}{cmyk}{0.57,0.3,0.0,0.0}

\usepackage{booktabs} % added by TK
\usepackage{multirow} % added by TK
\usepackage{adjustbox}
\usepackage{caption} %
\usepackage{orcidlink} 
\begin{document}

   \title{Fine-scale opposite-polarity magnetic fields in a solar plage revealed by integral field spectropolarimetry}

   \author{G. Liu
          \inst{1} \orcidlink{0009-0001-7029-5065}, I. Mili\'{c}\inst{1,2,3} \orcidlink{0000-0002-0189-5550}, 
          J.S. {Castellanos~Dur\'an} \inst{4} \orcidlink{0000-0003-4319-2009}, J.M. Borrero \inst{1} \orcidlink{0000-0003-4908-6186}, M. van Noort \inst{4}, C. Kuckein \inst{4,5,6} \orcidlink{0000-0002-3242-1497}
          }
   \institute{Institute for Solar Physics (KIS), Georges-K\"ohler-Allee 401A, 79110 Freiburg, Germany
   \and Faculty of Mathematics, University of Belgrade, Studentski Trg 16, 11000 Belgrade, Serbia \and Astronomical Observatory, Volgina 7, 11060 Belgrade, Serbia \and Max-Planck Institute f\"{u}r Sonnensystemforschung, Justus-von-Liebig-Weg 3, 37079 G\"{o}ttingen, Germany \and 
   Instituto de Astrof\'isica de Canarias (IAC), V\'ia L\'actea s/n, E-38205 La Laguna, Tenerife, Spain 
    \and
    Departamento de Astrof\'\i sica, Universidad de La Laguna, E-38206 La Laguna, Tenerife, Spain \\
   \email{gaojian.liu@email.uni-freiburg.de}}
  
   \date{Received ; accepted }

% \abstract{}{}{}{}{} 
% 5 {} token are mandatory
 
  \abstract
   {Plages are small concentrations of strong, nearly vertical magnetic fields in the solar photosphere that expand with height. A high spatial and spectral resolution that can resolve their fine structure is required to characterize them, and spectropolarimetric capabilities are needed to infer their magnetic fields.}
   {We constrain the 3D fine structure of the magnetic field in the photosphere of a solar plage from a unique spectropolarimetric dataset with a very high spatial and spectral resolution and a fast temporal cadence.}
   {We analyzed spectropolarimetric observations of a solar plage in the two magnetically sensitive spectral lines of neutral iron around 630\,nm. The observations were obtained with MiHI, which is an integral field unit attached to the Swedish Solar Telescope. MiHI obtained diffraction-limited, high-cadence observations with high spectral fidelity. These observations were interpreted using the spectropolarimetric inversion with magnetohydrostatic constraints, which allowed us to recover the magnetic and thermodynamic structure of the plage on a geometrical scale.}
   {The inversion results reveal that the magnetic field can reach up to 2\,kG and that it expands significantly from the deep to the mid-photosphere. Weaker ($\approx 200$\,G), and very small (subarcsecond) vertical magnetic loops lie beneath this canopy, rooted in the photosphere.}
   {This novel picture of a solar plage, in which weak opposite-polarity field patches surround the main polarity, provides new insight into convection in strongly magnetized plasma. 
   }
     \titlerunning{Small-scale opposite polarities in solar plage photosphere}
    \authorrunning{Liu et al.}
   \keywords{Sun: photosphere; Sun: faculae, plages; Sun: magnetic fields}

   \maketitle

%________________________________________________________________

\section{Introduction}
\label{sec:intro}

Solar plages are considered to be the roots of coronal loops. They exhibit strong, nearly vertical magnetic fields in the photosphere that expand and incline with height and eventually become more horizontal. They form the chromospheric canopy \citep[][]{VMP1997plage}. Their enhanced brightness in the UV is a crucial component in solar irradiance variations \citep{Krivova2003var} and a strong indicator of solar and stellar activity \citep{Chatzistergor2022plag}. Furthermore, plages are a potential conduit for energy transport from the photosphere upward because their vertical magnetic fields enable the propagation of waves that contribute to chromospheric \citep[][]{Morosin2022heating} and coronal heating  \citep{vanDoors2020review}. Historically, plages were modeled in the framework of the thin magnetic flux tube approximation  \citep{Spruit1976thintubes}, and they were characterized using spectropolarimetry to devise a canonical atmospheric model of these features  \citep{Solanki1992contbright}. A reconciliation of their structure and radiative properties is still somewhat of an open question, especially in the context of modeling observations away from the disk center, where plages are seen as faculae \citep[e.g.][]{Steiner2005fac, Albert2023A&A...678A.163A}.  

\cite{Buehler2015plagesp} analyzed plage observations obtained by the solar optical telescope spectropolarimeter on board the Hinode satellite \citep[SOT-SP,][]{Kosugi2007hinode} using spatially coupled spectropolarimetric inversions  \citep{vanNoort2012coupled, CastellanosDuran2024...modest}. They reported indications of opposite-polarity magnetic fields around the plage, but did not provide a temporal context. \cite{Morosin2020canopy} analyzed high-resolution observations in the \ion{Ca}{II}\,854.2\,nm line and focused on the canopy fields in the chromosphere. More recently,  \citet{Joao2023dkistplage} used the visible spectropolarimeter \citep[ViSP,][]{deWijn2022visp} at the 4m Daniel K. Inouye solar telescope \citep[DKIST,][]{Rimmele2020dkist} to constrain the properties of the magnetic field of the plage based on deep spectropolarimetry in the \ion{Fe}{I}\,630\,nm and \ion{Ca}{II}\,854.2\,nm lines. These observations offered a very high signal-to-noise ratio, but limited spatial resolution because the images could not be restored \citep[e.g.][]{vanNoort2017ImageRestoSpectra}. 

 \begin{figure*}[htbp]
    \centering
    \includegraphics[width=0.34\textwidth]{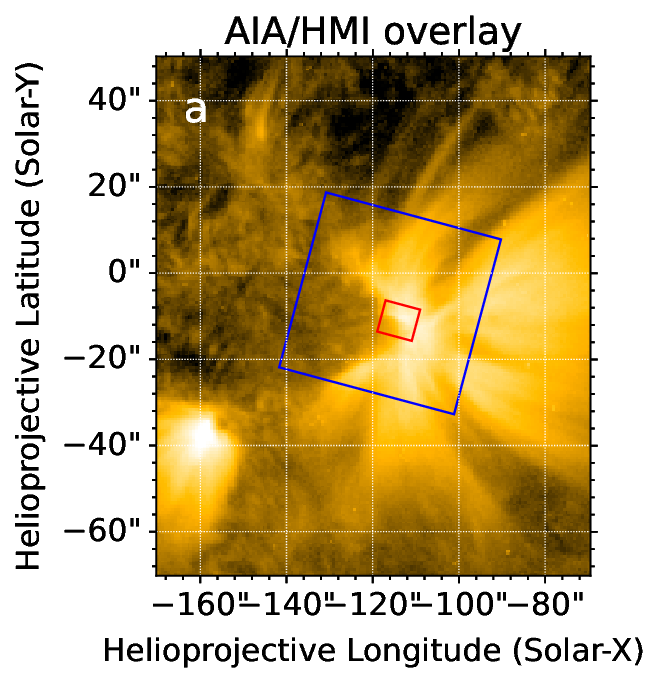}
    \includegraphics[width=0.65\textwidth]{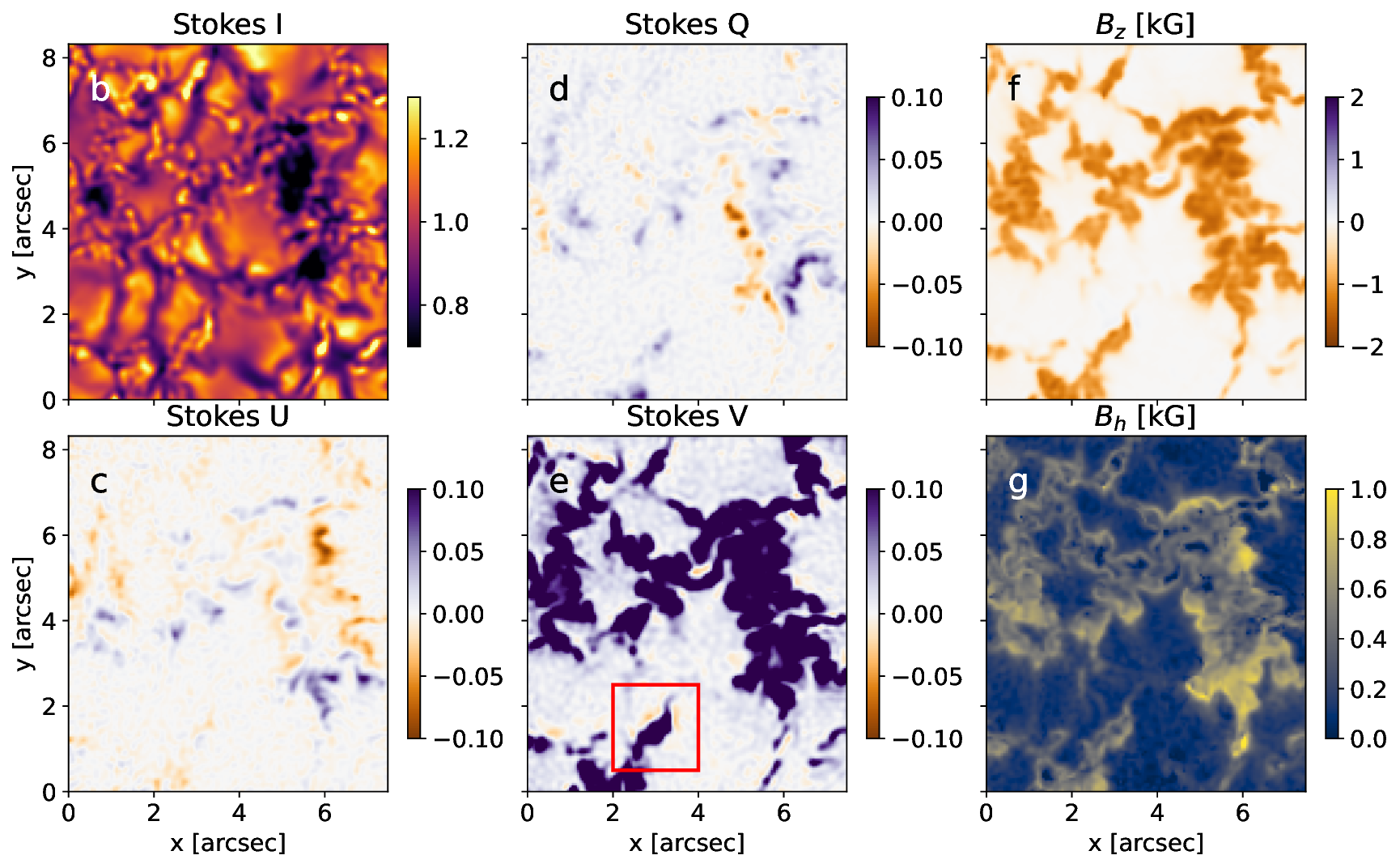}
    \caption{Left: AIA 17.1\,nm image at the start of the observing period, overlaid on an HMI magnetogram \citep{Schou2012}. The red square shows the MiHI field of view. Right: Maps of the four Stokes parameters, normalized to the quiet Sun at the disk center, and integrated over a small wavelength range in the continuum (Stokes $I$) center of the 630.2\,nm line (Stokes $Q,U$) and line wing (Stokes $V$), followed by the maps of the vertical ($B_z$) and horizontal ($B_h$) magnetic field inferred by pyMilne. The red square highlights the opposite polarities.}
    \label{fig:obs1}
\end{figure*}

Ideally, a high spatial resolution is required to characterize the plage structure so that the fine structure in the plane of the sky is resolved. A high spectral resolution is required to constrain the variations in the physical quantities along the line of sight (LOS), and spectropolarimetric capabilities are needed to infer the magnetic fields. A temporal cadence of some tens of seconds is desirable to study the evolution of the physical parameters on these small scales and to constrain the wave properties. An obvious candidate for conducting observations like this is an integral field unit (IFU), such as the Microlensed Hyperspectral Imager  \citep[MiHI,][]{vanNoort2022A&AMihiInstrument}. Only a prototype of this instrument currently exists for the visible wavelengths.
This prototype offers a limited field of view (FOV) that is instantaneously sampled with a high spectral and spatial resolution using a polarimeter. The data are spatially restored to a post facto selectable time cadence. During its commissioning and testing period at the Swedish Solar Telescope  \citep[SST,][]{Scharmer2003sst}, datasets were collected in the H$\alpha$, \ion{Na}{I}\,D1, and \ion{Fe}{I}\,630\,nm pair, some of which were of sufficient quality for scientific use \citep[see][]{vanderVoort2023A&AMiHIHalpha, Chae2024MiHi, Chitta2024MiHI}.

We analyze a spectropolarimetric dataset covering a solar plage near the disk center in the two photospheric magnetically sensitive neutral iron lines around 630\,nm. To quantitatively interpret the observed Stokes profiles and infer the variation in the physical parameters in 4D ($x,y,z,t$) in the observed atmospheric patch, we use the Milne-Eddington inversion code, pyMilne \citep{Jaime2019pm}, and a depth-stratified inversion code, FIRTEZ \citep{borrero_mhs_I}. FIRTEZ uses the magnetohydrostatic approximation to reconstruct the geometrical structure of the observed atmosphere. The high-resolution high-spectral fidelity imaging spectropolarimetry combined with a robust and physically motivated inversion approach results in a comprehensive picture of the plage that reveals several physical aspects that could not be probed in detail before. In this Letter, we report the detection of small-scale magnetic patches of opposite polarity around the plage that resemble very low-lying loops (at the base of the photosphere) that close over a distance of $\SI{200}{km}$. 

\section{Observations}
\label{sec:obs}

The data were taken at SST on August 9, 2018, from 09:35 to 10:00 UT at the position $-104.28'' {\rm E}, -11.33'' {\rm N}$. The field of view was sampled with $128\times115$ spatial points with a sampling of 0.065\,arcsec per pixel. This approximately corresponds to the critical sampling for the SST at this wavelength and amounts to 47\,km per pixel on the solar surface. The two \ion{Fe}{I} spectral lines were sampled from 629.906 to 630.494\,nm in steps of 1.0\,pm. The MiHI data are spectrally and spatially deconvolved as part of the reduction and restoration process. They achieve a spectral resolution higher than $3\times10^5$. The data were restored to a 10\,s temporal cadence for PyMilne inversions and 30\,s for FIRTEZ inversions (see below), which resulted in continuum noise levels of $1.6\times10^{-2}$ and $10^{-2}$. Although these numbers may appear to be nominally lower than those of more traditional observations, for example, Hinode SOT-SP, they are comparable to them or even better when the sampling and spatial and spectral point spread functions (PSF) of Hinode are taken into account, as discussed at length in  \cite{vanNoort2022MiHI3}.

The observations cover a part of a plage that hosts a brightening in the AIA\,17.1\,nm band \citep{Lemen2012}. It resembles the base of a large loop, but lacks a clear accompanying footpoint (see panel a of Fig.\,\ref{fig:obs1}). The MiHI FOV (panel b of Fig.\,\ref{fig:obs1}) contains a small-scale filigree pattern in the photosphere and several micropores. The Stokes $V$ in the wing of the \ion{Fe}{I}\,630.2\,nm line shows an almost exclusively unipolar signal even at these small scales ($<$ \SI{100}{km}). At several spatial locations near the filigree structures, however, Stokes $V$ changes sign compared to the plage field. This indicates opposite-polarity magnetic fields (see the red square in panel e of Fig.\,\ref{fig:obs1}). These opposite-polarity patches that surround the plage are the focus of our study. 

\begin{figure}[htbp]
    \centering
    \includegraphics[width=0.495\textwidth]{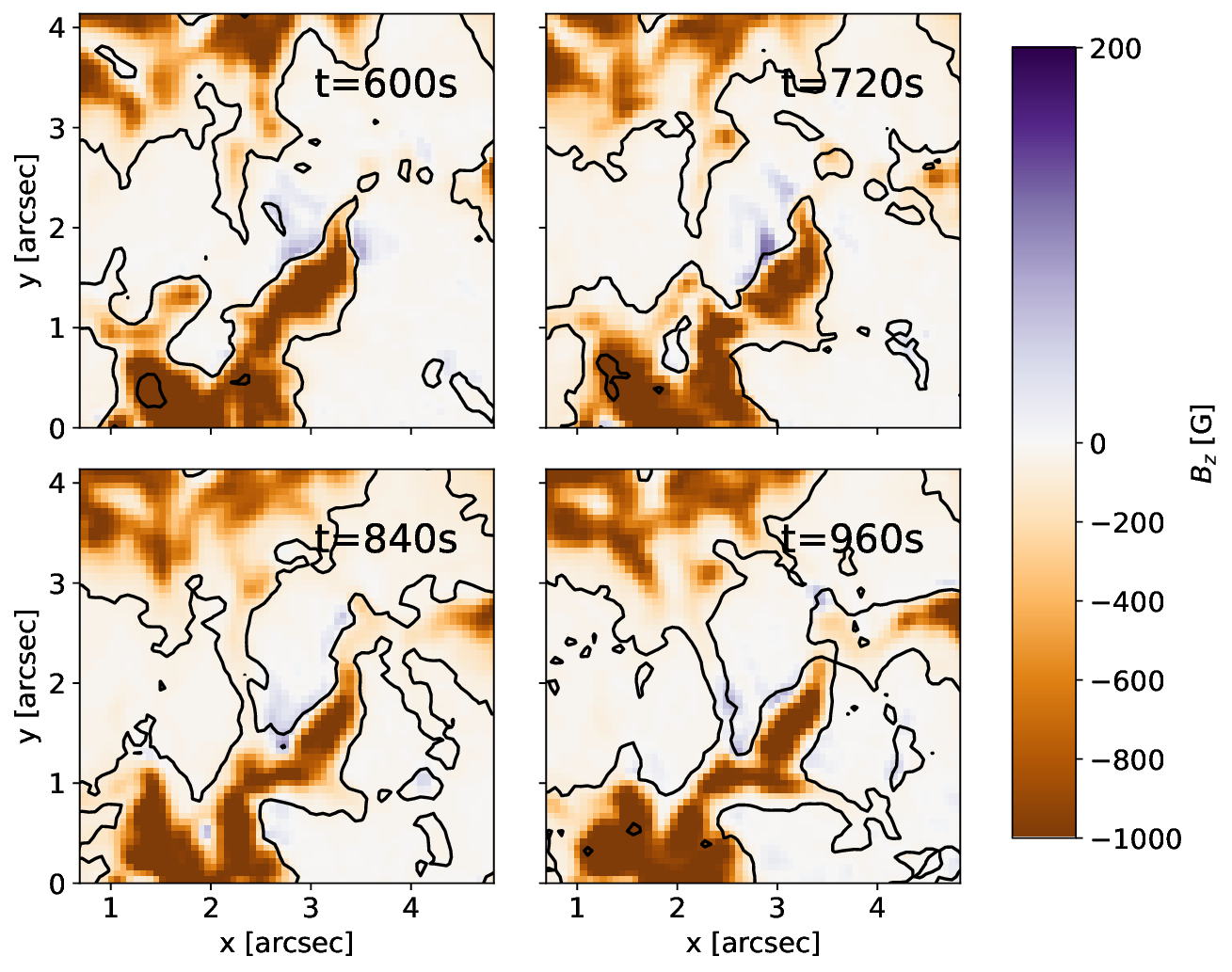}
    \caption{Magnetic field in the ROI retrieved by ME inversion in four instances of time separated by 120s. The times are given with respect to the start of the observations. The color scale is asymmetric. The contours mark areas with a horizontal magnetic field stronger than 200\,G.}  
    \label{fig:me_blos_ev}
\end{figure}

\section{Methods}
\label{sec:methods}

\subsection{Spatially regularized Milne-Eddington inversions}
\label{ssec:pm}
For a general estimate of the physical properties in the observed region and of their time evolution, we used PyMilne  \citep{Jaime2019pm}, an OpenMP parallelized spectropolarimetric inversion code based on the Milne-Eddington (ME) approximation. This code uses spatial coupling to account for the telescope PSF and spatial regularization to better interpret low-amplitude signals  \citep{Jaime2024spacetreg}. As discussed by  \citet{Orozco2010hinodelines} and \citet{borrero2014issi}, when these inversions are applied to the \ion{Fe}{I}\,630\,nm lines, they retrieve magnetic field values at a continuum optical depth of about $\log\tau_c = -1$ to $\log\tau_c = -1.5$, which corresponds to the mid-photosphere. As the MiHI data are spatially restored during the reduction process, no explicit spatial coupling through the telescope PSF is required in the inversion process. We only used spatial regularization, which helps us to interpret the signals close to the noise level of the data. As the Milne-Eddington inversion assumes a constant run of the magnetic field vector with depth, it will always retrieve the magnetic field in concordance with the polarity of the Stokes $V$ signal.

\subsection{Magnetohydrostatic inversions}
\label{ssec::firtez}

FIRTEZ is a depth-stratified spectropolarimetric inversion code that is formulated in the geometrical height scale  \citep{PastorYabar2019frzdz}. This code allowed us to take the variation in the physical parameters along the line of sight into account. The code uses a detailed equation of state for partially ionized gas and the magnetohydrostatic assumption  \citep{borrero_mhs_I} to relate the thermodynamic to the magnetic properties of the atmosphere and to infer an absolute common geometrical height scale for the entire field of view. Based on this, the Wilson depression in sunspots  \citep{Borrero2021mhsIII} can be inferred, for example. The FIRTEZ inversion is based on the concept of nodes, which are fixed-depth locations that control and constrain depth variations in physical quantities. The details of the inversion configuration are presented in Appendix\,\ref{app:frz_setup}.

\begin{figure}[htbp]
    \centering
    \includegraphics[width=0.495\textwidth]{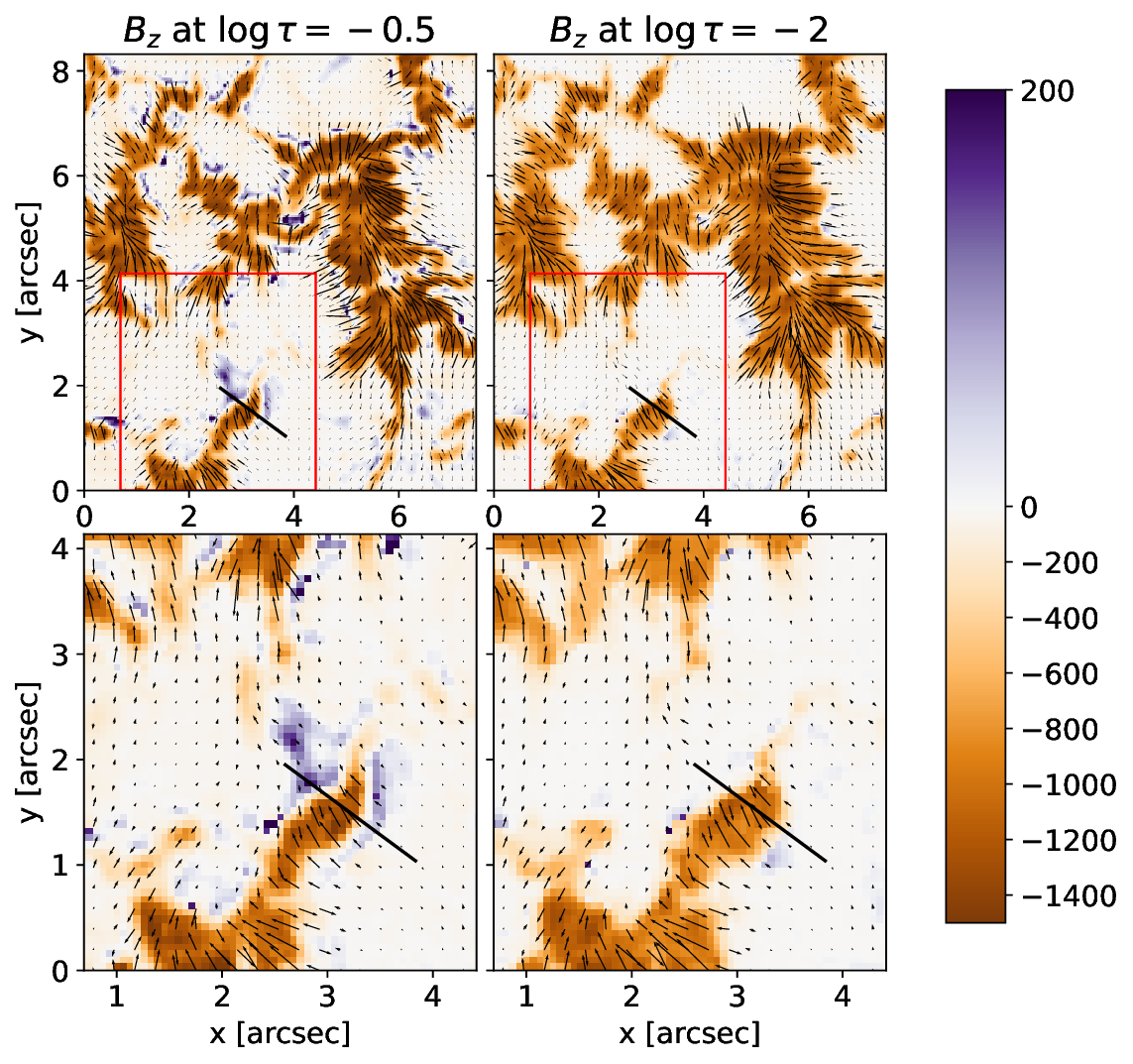}
    \caption{Magnetic field maps at the layer corresponding to $\log\tau=-0.5$ and $\log\tau=-2$. The color maps correspond to the vertical component and the arrows to the horizontal component of the magnetic field vector. The lower two panels show a zoom-in of the area denoted by the red rectangle in the upper two panels. The black line corresponds to the vertical slice shown in Fig.\,\ref{fig:vertical_cut}. The color-bar scale for $B_z$ is asymmetric.}
    \label{fig:bmaps}
\end{figure}

\section{Results and discussion}
\label{sec:results}

The pyMilne inversion provides general insight into the time dependence of the physical structure of the plage. The vertical field is almost exclusively unipolar, with a strength of up to 2\,kG (panel e of Fig.\,\ref{fig:obs1}). The horizontal field is strongest around the plage boundaries, where it reaches strengths of up to 1\,kG (panel f of Fig.\,\ref{fig:obs1}). A detailed view of the magnetic feature around $x=3'',y=2''$ (the region of interest; ROI) reveals patches of an opposite-polarity field (positive polarity field), with a vertical magnetic field strength of about $200$\,G (blue regions in Fig.\,\ref{fig:me_blos_ev}). As expected, they coincide with the locations in which the Stokes $V$ changes sign compared to the rest of the plage (see Fig.\,\ref{fig:stokes_fits} for examples of these Stokes $V$ profiles). The main (negative) and opposite (positive) polarities are separated by no more than 2 pixels, suggesting a loop-like structuring on scales of $\SI{200}{km}$ or below. The contours in Fig.\,\ref{fig:me_blos_ev} encircle locations in which the horizontal field is stronger than 200\,G. These locations also include the pixels that separate the positive and negative polarities, which further reinforces the idea of a loop geometry. These patches of opposite-polarity fields persist for the whole duration of the observations. ($\approx$25\,min), and their spatial structuring changes slightly on scales of several minutes. To our knowledge, these opposite polarities are directly detected in a plage region here for the first time, and they are shown to be stable on timescales comparable to or even exceeding the convection turnover time. 

\begin{figure*}[htbp]
    \centering
    \includegraphics[width=0.995\textwidth]{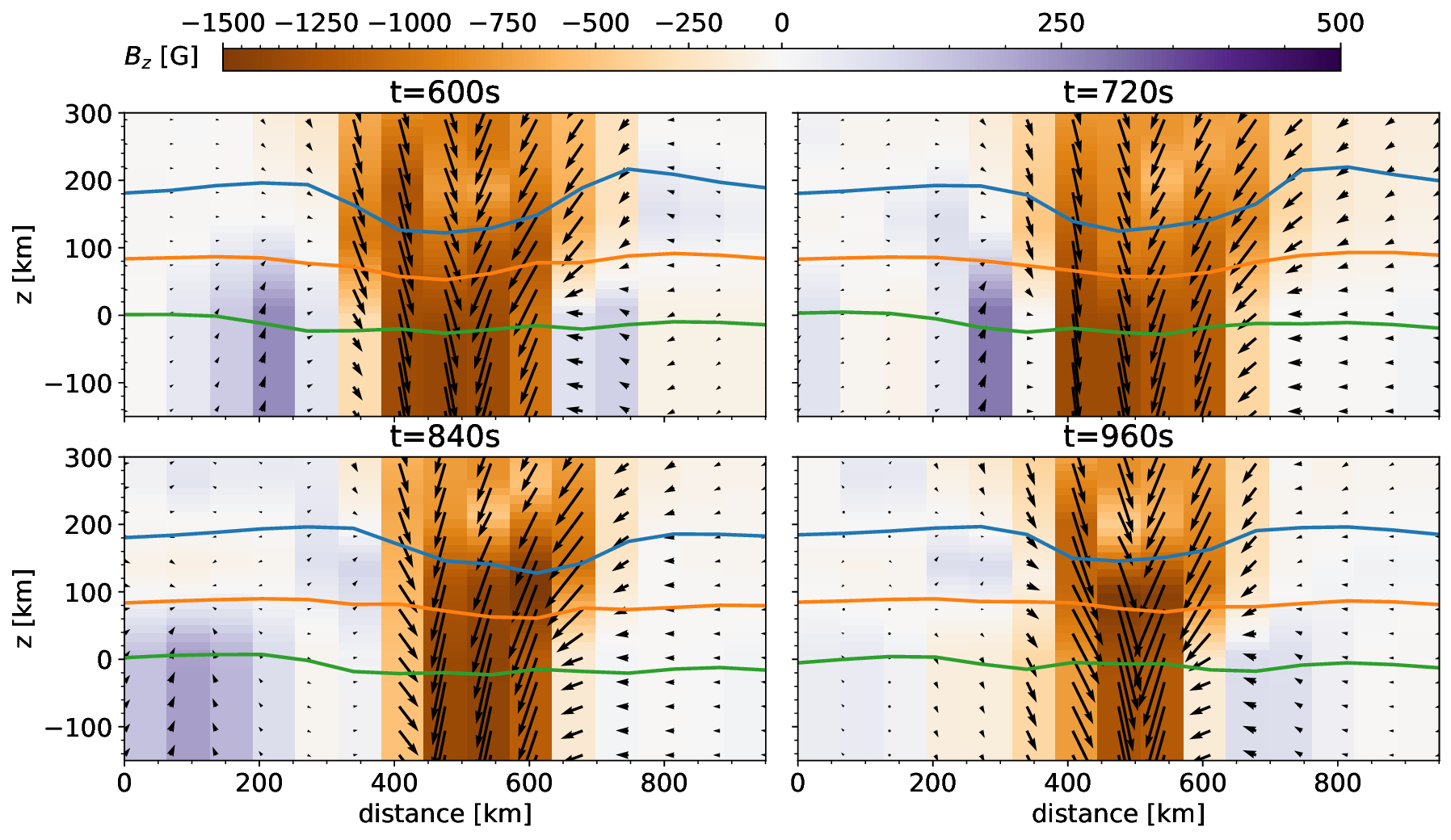}
    \caption{Cut showing the inferred magnetic field vector in the vertical plane that corresponds to the black line shown in Fig.\,\ref{fig:bmaps}. The green, orange, and blue lines denote $\log\tau=0,-1, and -2$ surfaces, respectively. The arrows indicate the direction and strength of the projection of the magnetic field vector on the vertical plane.}
    \label{fig:vertical_cut}
\end{figure*}

The FIRTEZ inversions are more informative about the depth stratification of the opposite-polarity patches. The two lower panels of Fig.\,\ref{fig:bmaps} show the magnetic field at $\log\tau=-0.5$ (left) and $\log\tau=-2$ (right) at the instance $t=\SI{600}{s}$ after the start of the observations. In the 3D model retrieved by the FIRTEZ code, these two optical depth surfaces are separated by 150\,km on average. The opposite polarities are visible at $\log\tau=-0.5$ and largely disappear at $\log\tau=-2$. These results further support the picture of low-lying small-scale loops that close before they extend to the mid-photosphere. The sizes of these loops would be similar to those found in the quiet Sun using Hinode SOT-SP and the Dutch Open telescope  \citep{Marian2009loopshinode}. However, the spatial context is drastically different (quiet Sun versus a plage), and so is the temporal evolution: The opposite polarities we found lived for tens of minutes. If the origin of the opposite polarity found in our data were the emergence of $\Omega$-shaped loops, then this process stops in the photosphere in our case.

Furthermore, the map at $\log\tau=-0.5$ reveals ubiquitous opposite polarities around the plage, in sharp contrast with the results obtained by the Milne-Eddington inversion, which only showed sporadic patches. Because the spectral lines are less sensitive to the magnetic field at the base of the photosphere ($\log\tau=-0.5$), it cannot be excluded that they are artifacts generated by the node interpolation used in the inversion code. The clear spatial structure of the magnetic field at this layer, retrieved by FIRTEZ, and the temporal consistency of the result probably mean that these field patches are the most consistent explanation for the observations. 

Fig.\,\ref{fig:vertical_cut} shows a vertical cut over the ROI, along the black line in Fig.\,\ref{fig:bmaps}, and it provides a more detailed insight into the topology of the magnetic field of the plage and its surroundings. The expansion of the plage field is evident and substantial even for this small height range. The structure of the magnetic feature is asymmetric, possibly as a result of the interaction of the plage field with convective motions around it. Figs.\,\ref{fig:me_blos_ev} and \ref{fig:vertical_cut} reveal that the shape and height stratification of these opposite-polarity patches change with time, but a certain amount of opposite polarity can be detected throughout the 25\,min of the observations. The opposite polarities appear to form loop-like structures at the sides of the plage that close over distances of approximately 200-300\,km. These structures resemble a canopy that is structured on much smaller scales than the chromospheric canopy. Additional insight into the time evolution of these loops would be provided by the inference of the horizontal velocities by either employing high-resolution optical flow tracking  \citep[e.g.][]{AAR2017deepvel} or by modeling the induction equation  \citep{Schuck2005dave}. 

\section{Conclusions}
\label{sec:conclusions}

We presented observations and a first analysis of very high-resolution spectropolarimetric IFU observations of a part of a solar plage near the disk center using the pair of magnetically sensitive neutral iron lines around 630\,nm. The MiHI prototype mounted at the SST provided data with a high spatial resolution, high spectral fidelity, and fast cadence. Combined with the state-of-the-art inversion codes PyMilne and FIRTEZ, the data provide an unprecedented insight into the spatio-temporal structure of the plage and surrounding convection. Our results show the clear signature of weak small-scale opposite-polarity field patches around the plage that persist over timescales of minutes or longer. This exceeds the convection turnover time in the solar photosphere. These opposite-polarity field patches are detectable directly from the signs of the Stokes $V$ profiles, and the strength of the vertical field can be inferred from simple but robust Milne-Eddington inversions. A more detailed depth-stratified inversion on a geometrical height scale using the FIRTEZ spectropolarimetric code with magnetohydrostatic constraints revealed significant variations in the magnetic field with height and opposite polarities at $\log\tau=-0.5$ in multiple areas around the plage. This finding will be further confirmed through a detailed comparison of the inverted atmospheres with simulations of the surface convection using the code MURaM  \citep{Vogler2005muram, Przybylski2022muramc}. The multiline high-resolution data collected by the third flight of the SUNRISE  \citep{Korpi-Lagg2025SunriseIII} balloon-borne telescope, the FISS-SP instrument at GST \citep{Cao2010GST}, and TRIPPEL-SP at SST \citep{Saranthan2021stray} can also provide larger-scale insight into these magnetic features. 

The origin of these opposite-polarity patches around the plage is not clear. The patches are universally situated next to the main polarity, suggesting that the plage field is tangled through vigorous convection. This process would have to be ubiquitous and persistent to explain the considerable lifetimes of the opposite polarities. The emergence of small-scale loops, similar to the quiet Sun, would be another explanation. It is unclear why this emergence would prefer an orientation in which the loop footpoint that has the same polarity as the plage footpoint always faces the plage. Another potential scenario is that granule-sized loops that form a low-lying canopy were compressed through the convective motions to form these micro-canopies. Although our time sequence covered several granular timescales, our preliminary analysis did not reveal an event like this. \citet{Chitta2019sstplage} also found small-scale opposite polarities around a plage in high-resolution SST observations. The authors identified them as important for the energy transfer toward the corona. In contrast to their findings, the opposite-polarity patches we reported are much smaller and show no obvious emergence. Furthermore, our dataset has a higher spectral fidelity, so we were able to infer a detailed 3D magnetic field configuration, which does not seem to suggest cancellation events. 

In addition to presenting new insights into the photospheric magnetism of a solar plage, we showed the diagnostic potential of integral field spectropolarimetry at the diffraction limit, combined with physically motivated inversion approaches. This combination provides the best possible use of high-resolution spectropolarimetric observations of the Sun and an unprecedented view of the interaction of plasma, magnetic field, and radiation. The development of such instruments and diagnostic tools is paramount for making the best use of the existing and upcoming large solar telescopes such as DKIST \citep{Rimmele2020dkist}, WeHot, EST \citep{cqn2022est}, and NLST.

\begin{acknowledgement}
We gratefully acknowledge stimulating discussions with O. Steiner, Vigeesh G., D. Przybylski, and P. K\"{a}pyl\"{a}. IM acknowledges the financial support from the Serbian Ministry of Science and Technology through the grants 451-03-136/2025-03/200104 and 451-03-136/2025-03/200002. CK acknowledges grant RYC2022-037660-I funded by MCIN/AEI/10.13039/501100011033 and by "ESF Investing in your future". 
This research has made use of NASA's Astrophysics Data System.  
\end{acknowledgement}
\bibliography{plage}

\begin{appendix}
\label{app:opposite_v}

\section{Example Stokes profiles}

\begin{figure}[htbp]
    \centering
    \includegraphics[width=0.495\textwidth]{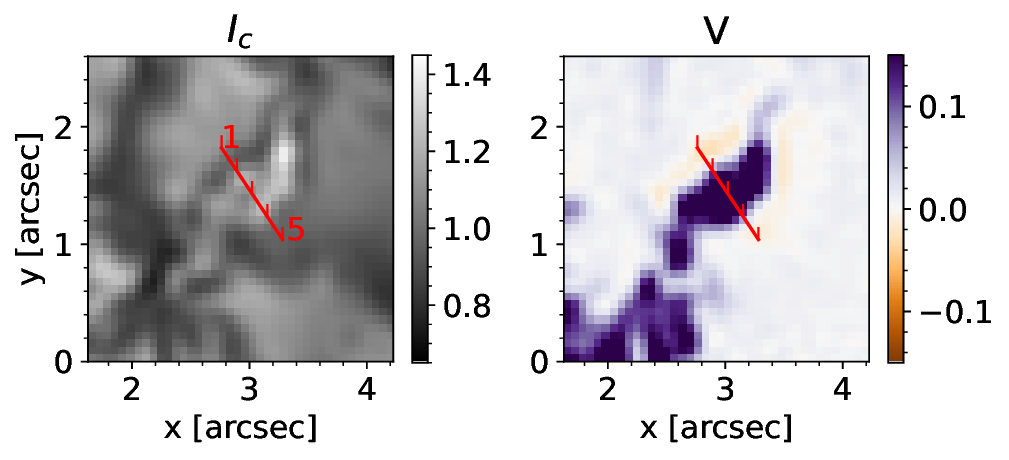} \\
    \includegraphics[width=0.495\textwidth]{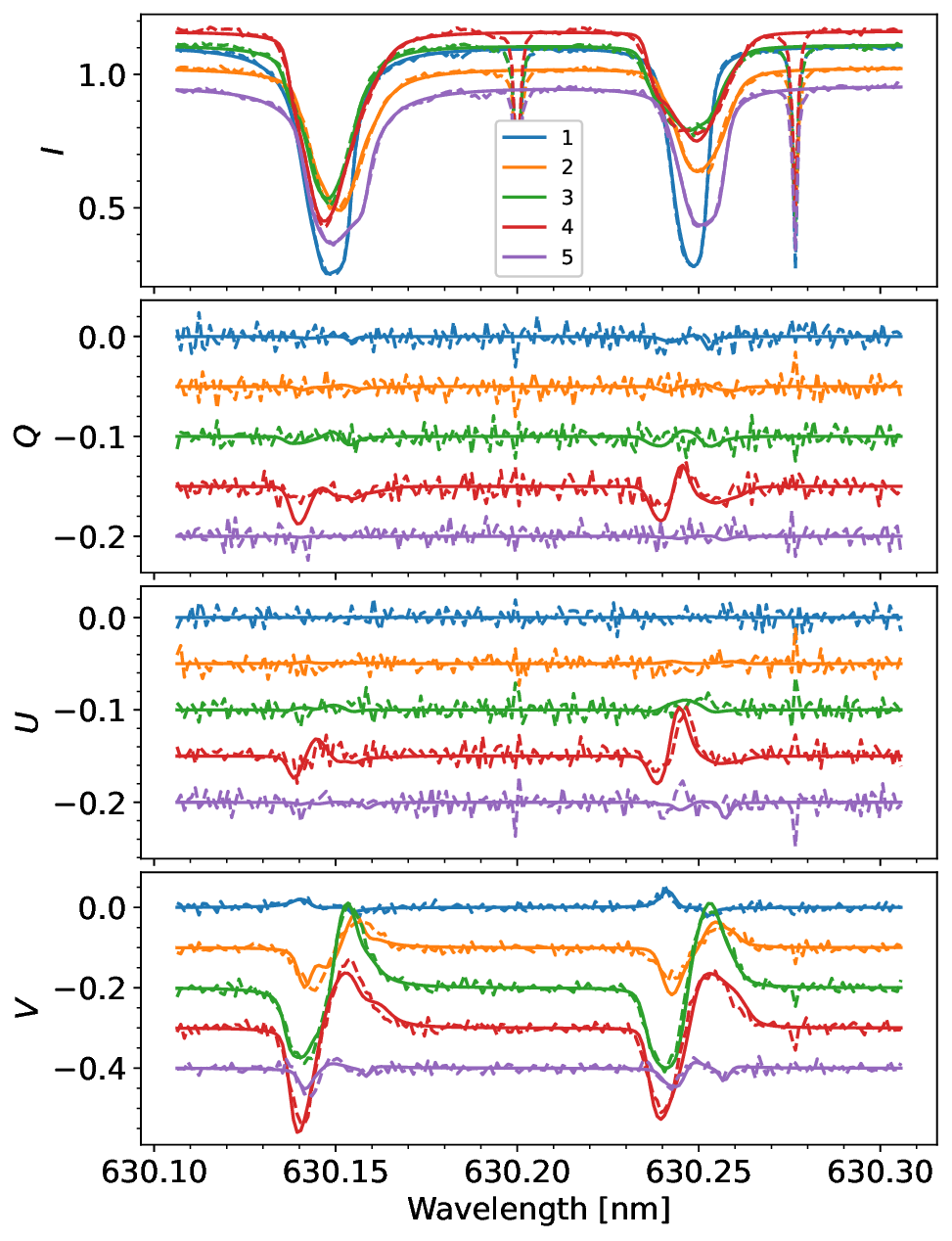}
    \caption{Top: Continuum intensity (left) and the Stokes $V$ in the line wing (right) in the ROI at the time $t=600$\,s with respect to the start of the observations. Bottom four panels: Stokes $I,Q,U,V$ at the five selected locations. Stokes $Q,U,V$ have been offset vertically for clarity. Dashed and full lines denote observed and FIRTEZ-fitted spectra. All Stokes components are normalized to the continuum at the quiet Sun at the disk center.}
    \label{fig:stokes_fits}
\end{figure}

While the Stokes $V$ map in Fig.\,\ref{fig:obs1} shows, upon a closer inspection, the opposite polarities in the observations, we showcase different types of Stokes profiles found in our data in Fig.\,\ref{fig:stokes_fits}. Pixels 2 and 4 are situated at the plage boundary: their asymmetry suggests strong velocity gradients, as expected in plage. Interestingly, only the pixel 4 exhibits strong linear polarization. For the scope of this paper, the most interesting are pixels 1 and 5, which have the opposite polarity compared to the other three. Pixel 5, specifically, shows a three-lobed Stokes $V$ profile that indicates velocity and magnetic field gradient with the change of polarity with depth. We also want to draw attention to the variety of the Stokes profiles observed over such a small spatial domain. All the fits in Fig.\,\ref{fig:stokes_fits} are obtained with the FIRTEZ code.

\begin{figure}[h]
    \centering
    \includegraphics[width=0.495\textwidth]{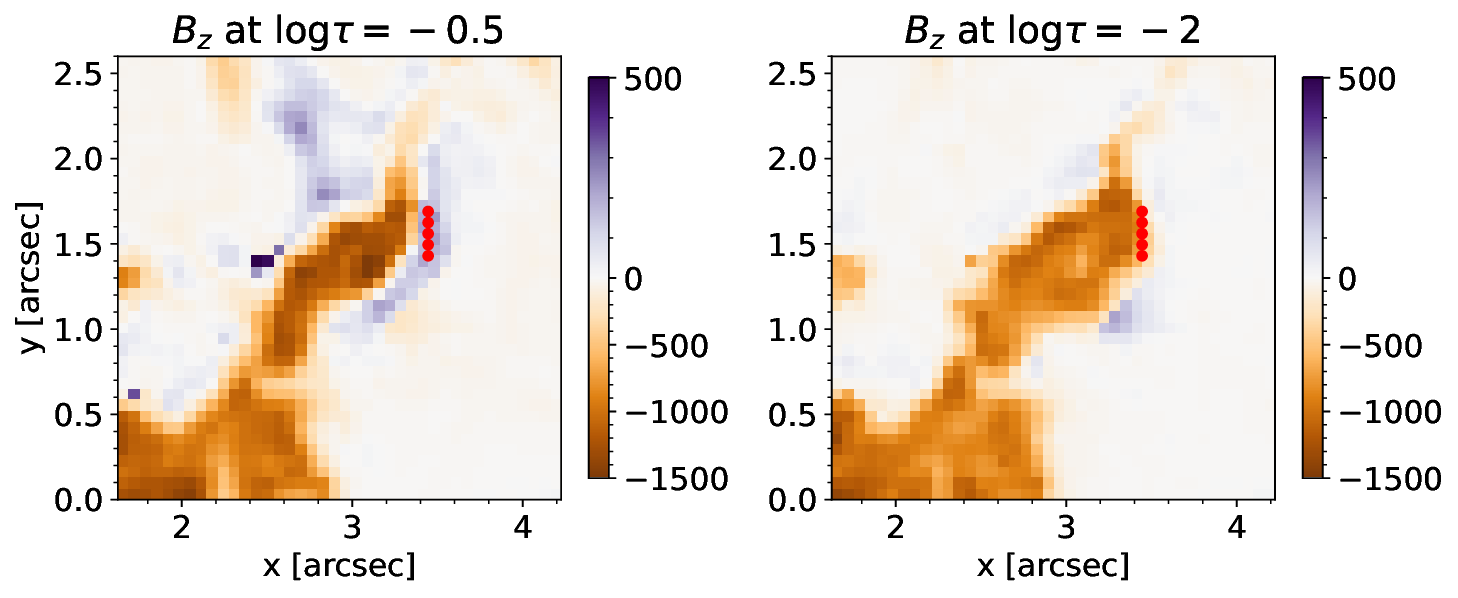}
    \includegraphics[width=0.495\textwidth]{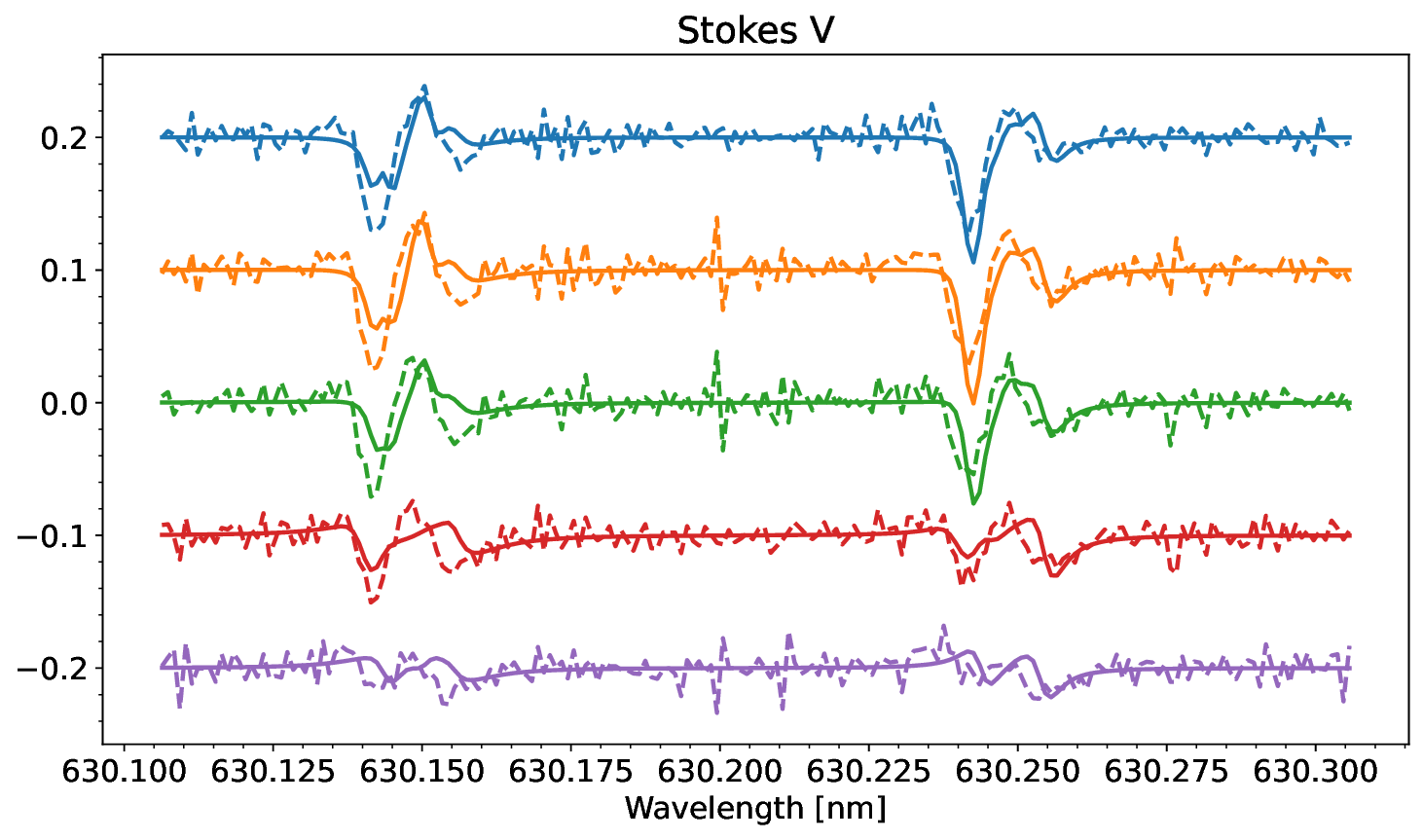}
    \caption{Top: Vertical magnetic field maps at the layer corresponding to $\log\tau=-0.5$ and $\log\tau=-2$. Note the asymmetric colorbar scale. Bottom: Stokes $V$ profiles corresponding to the five red points in the top panels. The profiles have been offset vertically for clarity. Dashed and full lines denote observed and FIRTEZ-fitted spectra.}
    \label{fig:stokes_gradients}
\end{figure}

Additionally, we showcase the Stokes $V$ profiles in several places where the magnetic field inferred by FIRTEZ changes polarity with height. Example pixels are marked with red points in the upper panel of Fig.\,\ref{fig:stokes_gradients}. They exhibit opposite polarity in the lower parts of the atmosphere ($\log\tau=-0.5$), and no opposite polarity in the upper parts of the atmosphere ($\log\tau=-2$). Fig.\,\ref{fig:stokes_gradients} shows the normalized $V$ profiles in the selected pixels. Each profile is complex and shows multi-lobed, asymmetric behavior, indicative of the strong gradients of magnetic field and velocity. While the exact atmosphere structure inferred by our inversions is subject to noise effects and parameter degeneracies, the shape of these profiles indicates depth-dependence of the magnetic field.

\section{FIRTEZ setup}
    \label{app:frz_setup}

We initialized the magnetic field for FIRTEZ inversions using results from PyMilne and used a multi-cycle approach where different physical parameters have different numbers of nodes in each cycle and different Stokes parameters are weighted differently. We use equal weights for all Stokes parameters in the first cycle and four times larger weights for Stokes $V$ in the second cycle. The first cycle uses $(6,2,1,2)$ nodes for the temperature, LOS velocity, and horizontal and vertical field components, respectively, while the second cycle uses $4,4,2,4$ nodes. The first cycle attempts to constrain the temperature structure, while the second cycle polishes out the details in the velocity stratification and focuses on reproducing Stokes $V$ profiles. The map-averaged $\chi^2$ of the obtained fits, given the noise level of $10^{-2}$ in the units of the quiet Sun continuum, is around unity, roughly indicating that the model complexity is appropriate for the data in question (see Fig.\,\ref{fig:stokes_fits}, for some examples of fitted Stokes parameters). Strictly speaking, even the image-restored data will still have some residual stray light - a contribution of the uncorrected, wide wings of the spatial PSF of the optical system, to each of the pixels in the image  \citep[see][for a more in-depth discussion]{Saranthan2021stray}. The amount of stray light can have a substantial effect on the physical structure of the inverted atmosphere, most notably on the inferred velocities and magnetic field. We attempted to correct the stray light by manually subtracting different amounts ($0-30\%$) of the mean polarized spectrum over the FoV from the data and inverting the data again. Interestingly, results for stray light levels for ($0-10\%$) were fairly consistent, while the higher levels of stray light resulted in worse fits (higher $\chi^2$ values). An independent test using SIR code  \citep{ruizcobo1992sir}, where the stray light level was set as a free parameter, confirmed this finding. This, again, led us to conclude that our data has excellent spatial resolution, which is especially important for probing small-scale phenomena.

After the satisfactory fit is obtained, the magnetic field is disambiguated using the approach of  \citet{Georgoulis2005disam}, and then the height stratification is adjusted to conform to MHS constraints  \citep{borrero_mhs_I}. This approach constrains the depth dependence of the magnetic field and enables the inference of the depression of the iso-$\log\tau$ surfaces due to the influence of the magnetic field. The final product is the 4D cube ($t,x,y,z$) of physical parameters corresponding to the series of input datacubes of Stokes parameters. Thanks to the MHS capabilities of FIRTEZ, we can investigate the detailed 3D structure without relying on the optical depth scale, which is a significant step forward compared to the other inversion codes.

\end{appendix}

%-----------------------------------------------------------

\end{document}